\documentclass{article}
\usepackage[utf8]{inputenc}
\usepackage{amsmath}
\usepackage{amsfonts}
\usepackage{amssymb}
\usepackage{graphicx}

\begin{document}

\title{On complex dynamics from reversible cellular automata}
\author{Juan Carlos Seck-Tuoh-Mora*, Genaro J. Martinez, \\
Norberto Hernandez-Romero, Joselito Medina-Marin, \\
Irving Barragan-Vite \\
AAI-ICBI-UAEH. Carr Pachuca-Tulancingo Km 4.5. \\
Pachuca 42184 Hidalgo. Mexico \\
Unconventional Computing Centre, University of the \\
West of England, BS16 1QY Bristol, United Kingdom \\
Escuela Superior de Computo, Instituto Politecnico \\
Nacional, Mexico \\
}
\date{September 2020}


\maketitle

\begin{abstract}

Complexity has been a recurrent research topic in cellular automata because they represent systems where complex behaviors emerge from simple local interactions. A significant amount of previous research has been conducted proposing instances of complex cellular automata; however, most of the proposed methods are based on a careful search or a meticulous construction of evolution rules. 

This paper presents the emergence of complex behaviors based on reversible cellular automata. In particular, this paper shows that reversible cellular automata represent an adequate framework to obtain complex behaviors adding only new random states. 

Experimental results show that complexity can be obtained from reversible cellular automata appending a proportion of about two times more states at random than the original number of states in the reversible automaton. Thus, it is possible to obtain complex cellular automata with dozens of states.  Complexity appears to be commonly obtained from reversible cellular automata, and using other operations such as permutations of states or row and column permutations in the evolution rule. The relevance of this paper is to present that reversibility can be a useful structure to implement complex behaviors in cellular automata. \\

\end{abstract}


Submitted to: Comm Nonlinear Sci Numer Simulat

\section{Introduction}

Reversible cellular automata (RCAs) have received much attention in the last years due to their information conserving property, which offers a framework to analyze interesting dynamics.  An RCA is defined by two evolution rules which are reversible each other yielding an invertible global dynamics. RCAs have many possible uses to implement and analyze universal systems, expansiveness or conservation laws \cite{kari-2018} \cite{martinez-2018} \cite{morita-2012}. However, up to our knowledge, they have not been applied to provide a systematic method to obtain complex cellular automata.

Rule 110 is a classic example of a complex cellular automaton; the evolution rule can generate mobile particles  (called self-localizations or gliders as well) in a periodic background. Thus, one way to obtain complexity is to define evolution rules producing periodic backgrounds and the interaction of particles \cite{Eppstein-2019}. Wolfram presented the best-known reference presenting a classification of complexity in elementary cellular automata in \cite{Wolfram84}. Since this seminal paper, other classifications were proposed; for instance, using the complexity of languages, equicontinuity and attractors \cite{Kurka-97}. Previous works have introduced instances of complex cellular automata with different characteristics (dimension, number of states, neighborhood size, and topology). However, most of these methods are based on a careful search or construction of evolution rules, and little attention has been paid to the systematic use of RCAs for this purpose. 

The present paper presents the application of RCAs for obtaining complexity in cellular automata. The desired behavior is achieved adding random states to the original reversible evolution rule, other operations conserving reversibility such as permutation of states and positions in the evolution rule are also investigated. This process can generate extended cellular automata with periodic backgrounds and interacting particles.


\section{Related work}


Much research in recent years has focused on the study of complexity in cellular automata. Theoretical studies include the use of several techniques to detect and classify complexity in cellular automata. For instance, tools from symbolic dynamics such as shift equivalence  \cite{Ruivo-2018}, subshifts of finite type \cite{Dennunzio-Formenti-Manzoni-2014} and topological entropy \cite{Zhang-Zhang-2015}  \cite{Xu-Li-Chen-Jin-2018} \cite{Guan-Chen-2015}  \cite{Pei-Han-Liu-Tang-Huang-2014}. Other techniques have also proposed, in particular, decision algorithms \cite{Autran-Formenti-2018}, measuring their ability to store and process information by particles \cite{Zhang-2018} and probability of words \cite{Boyer-Delacourt-Poupet-Sablik-Theyssier-2015}. 

Chaotic dynamical behaviors in the sense of topological entropy are considered in \cite{Chang-Akin-2016} for invertible one-dimensional linear cellular automata with several states. This work implies that, depending on the selected measurement, reversible cellular automata can generate interesting dynamical behaviors. 

Some of the current applications of complex cellular automata are devoted to simulate universal structures  \cite{Goles-Pedro-Theyssier-2018},  emulate systems engineering with different levels of complexity \cite{Nichele-Mathias-Risi-Tufte-2018} and model various cooperative strategies for the generation of new knowledge \cite{Jaeger-2018}. 

A primary current focus of research is how to utilize different types of cellular automata to yield complexity; for instance,  polynomial cellular automata \cite{Stone-18},  majority rules \cite{Garcia-Morales-2018},  Turing-universal cellular automata with prime and composite rules \cite{Riedel-Zenil-2018} and cellular automata with evolution rules defined with Kolmogorov complexity \cite{Peled-Carmi-2018}. A related work employed infinite Petri nets to simulate the elementary cellular automaton Rule 110 \cite{Zaitsev-2018}.

Given the relevance of complexity in cellular automata, some different techniques and approaches have been employed to study and characterize complexity. Hanson and Crutchfield propose domain filters to locate and classify particles in elementary cellular automaton rule 54 \cite{Hanson-Crutchfield-97}. The study of complex cellular automata as a tilling problem has been applied for the analysis and extension of topological properties in several dimensions \cite{Dennunzio-EtAl-14}. The asymptotic behavior of cellular automata in higher dimension according to a Bernoulli probability measure is studied in \cite{Delacourt-Hellouin-2017}, showing that cellular automata have the same variety and complexity of conventional Turing machines when self-organization emerges from random configurations. The limit behavior of cellular automata in an infinite time evolution is computed and classified by an automatic method based on finite automata and regular expressions in \cite{Oliveira-Ruivo-Costa-2016} and \cite{Ruivo-Oliveira-2017}. About techniques to identify particles directly in a periodic background, the $Z$-parameter has also been employed to find gliders in cellular automata \cite{wuensche1999classifying}; other techniques also include adding memory to the evolution rule \cite{martinez2013designing}, the use of genetic algorithms \cite{das1994genetic} \cite{sapin2003research} and the application of de Bruijn diagrams \cite{martinez2006gliders} \cite{martinez2008representation} \cite{martinez2008determining}.

Entropy and density of states are common measurements used to detect complex dynamics in cellular automata, due to these are easily defined and low time-consuming to be calculated. 

Shannon entropy complemented with Kolmogorov complexity has been employed to quantify the structural characteristics, in two-dimensional multi-state cellular automata \cite{Javid-Blackwell-Zimmer-Rifaie-2016} and periodic Coven cellular automata in \cite{Liu-Ma-2016}. The entropy of fractal type cellular automata is utilized in \cite{Ganikhodjaev-Saburov-Mohamad-2016}. An extension of Lempel-Ziv complexity measure to estimate the entropy density from random configurations has been addressed in \cite{Estevez-Lora-Nunes-Aragon-2015}. The entropy of the stochastic Fukui-Ishibashi traffic model was resolved in \cite{Salcido-Hernandez-Carreon-2018} obtaining exact analytical solutions. Entropy variations are measured in  \cite{Borriello-Walker-2017} to propose an information-based classification. Entropy is compared with the local structure theory to characterize density for elementary cellular automata in \cite{Fuks-2014}, illustrating the case of rule 26. A mean-field approximation for density has been investigated in \cite{Mendonca-2018} using elementary cellular automata in nonoverlapping generations both with crowded and dispersal neighborhoods. Mean-field approximation of density and Monte Carlo simulations were employed in probabilistic cellular automata to prove that they converge to a specific measure \cite{Ramos-Leite-2017}. Second-order phase transition in asynchronous cellular automata has been analyzed in \cite{Fuks-Fates-2015} using local structure theory of density to predict a qualitative change from an active phase into a stationary state with fluctuations. In \cite{adamatzky2006phenomenology} density in elementary cellular automata has been employed to find particles and complex behaviors.  In \cite{Seck-Medina-Martinez-Hernandez-2014} density in elementary cellular automata has been used to define interpolation surfaces and classify periodic, chaotic and complex behaviors. 

These articles confirm that entropy and density are currently applied in recent works for analyzing complexity. Therefore this paper uses these tools to detect complex cellular automata.

\section{Definition of Cellular Automata}


A one-dimensional cellular automaton $\mathcal{A}$ is an array of cells whose dynamics is locally defined by a finite set of states $S$ and mapping (or evolution rule) $\varphi: S^m \rightarrow S$, where $m$ is the neighborhood size of $\mathcal{A}$. Thus, a cellular automaton can be described as a tuple $\mathcal{A}=\{S, m, \varphi \}$. 

A global state (or configuration) of $\mathcal{A}$ is specified by $c: \mathbb{Z} \rightarrow S$ which assigns to every cell a state of $S$. Every cell $c_i$ has associated a neighborhood vector $N_i=\{i,n_1 \ldots n_{m-1}\}$ with $m$ relative positions (including $i$) specifying the neighbors of $c_i$. The evolution rule is applied over every $c_{N_i}$ to obtain the new state of $c_i$ and a new configuration $c'$. In this investigation, periodic boundary conditions are considered in the numerical simulations.

Definition of the evolution rule can be extended for sequences of states larger than the neighborhood size $m$. For an integer $m' > m$ and each sequence $w \in S^{m'}$, the evolution rule $\varphi$ can be applied over every complete neighborhood of $w$. Notice that there are $m'-m+1$ overlapping neighborhoods in $w$, then $\varphi(w)=w' \in S^{m'-m+1}$. Thus, $\varphi:S^{m'} \rightarrow S^{m'-m+1}$.

This feature can be used to simulate the original cellular automaton with another of neighborhood size $2$ and a larger number of states. If we take sequences of length $m'=2m-2$, then the evolution rule can be applied as $\varphi: S^{m'} \rightarrow S^{m-1}$; this means that $\varphi$ yields a mapping from two blocks of length $m-1$ into one block of the same size. Taking now a new set of states $K$ such that $|K| = |S^{m-1}|$ and there is a bijection from $K$ to $S^{m-1}$, then local dynamics of $\varphi$ can be simulated by an analogous evolution rule $\varphi':K^2 \rightarrow K$.

This simulation shows that every one-dimensional cellular automaton $\mathcal{A}=\{S, m, \varphi \}$ can be simulated by another $\mathcal{A'}=\{K, 2, \varphi' \}$, and we only have to study cellular automata of neighborhood size $2$ to understand the other cases.

In this case, the evolution rule $\varphi$ can be represented by a matrix $M$ where rows and columns are states in $S$ and for every $a,b \in S$, $M(a,b)=\varphi(ab)$.

\section{Parameters to classify complex cellular automata}

Rule 110 is composed of $2$ states and neighborhood size $3$; the evolution table is defined by the binary representation of $110$ as follows:

\[
\begin{array}{|ccc|c}
0&0&0&0 \\
0&0&1&1 \\
0&1&0&1 \\
0&1&1&1 \\
1&0&0&0 \\
1&0&1&1 \\
1&1&0&1 \\
1&1&1&0 \\
\end{array}
\]

Blocks of $4$ cells evolve into blocks of $2$ states; for instance, the block $0000$ evolves into $00$, $0001$  into $01$ and so on. Every block of two states can be identified with a number from $1$ to $4$. Hence Rule 110 can be represented with another cellular automaton of $4$ states and neighborhood size $2$. The corresponding evolution rule is illustrated by a table $M$ where the row and column indexes are neighbors, and every entry is the evolution of every neighborhood. In the following, every evolution rule $\varphi$ is represented by $M_\varphi$.  

\begin{equation} 
M_{110}=\begin{array}{c|cccc}
\multicolumn{1}{c}{}& 1 & 2 & 3 & 4 \\
\cline{2-5}
1& 1 & 2 & 4 & 4 \\ 
2& 3 & 4 & 4 & 3 \\ 
3& 1 & 2 & 4 & 4 \\ 
4& 3 & 4 & 2 & 1
\end{array}  
\label{Regla110-4h}
\end{equation}



\begin{figure}[th]
	\centering
	\includegraphics[width=1 \linewidth, height=0.65 \linewidth]{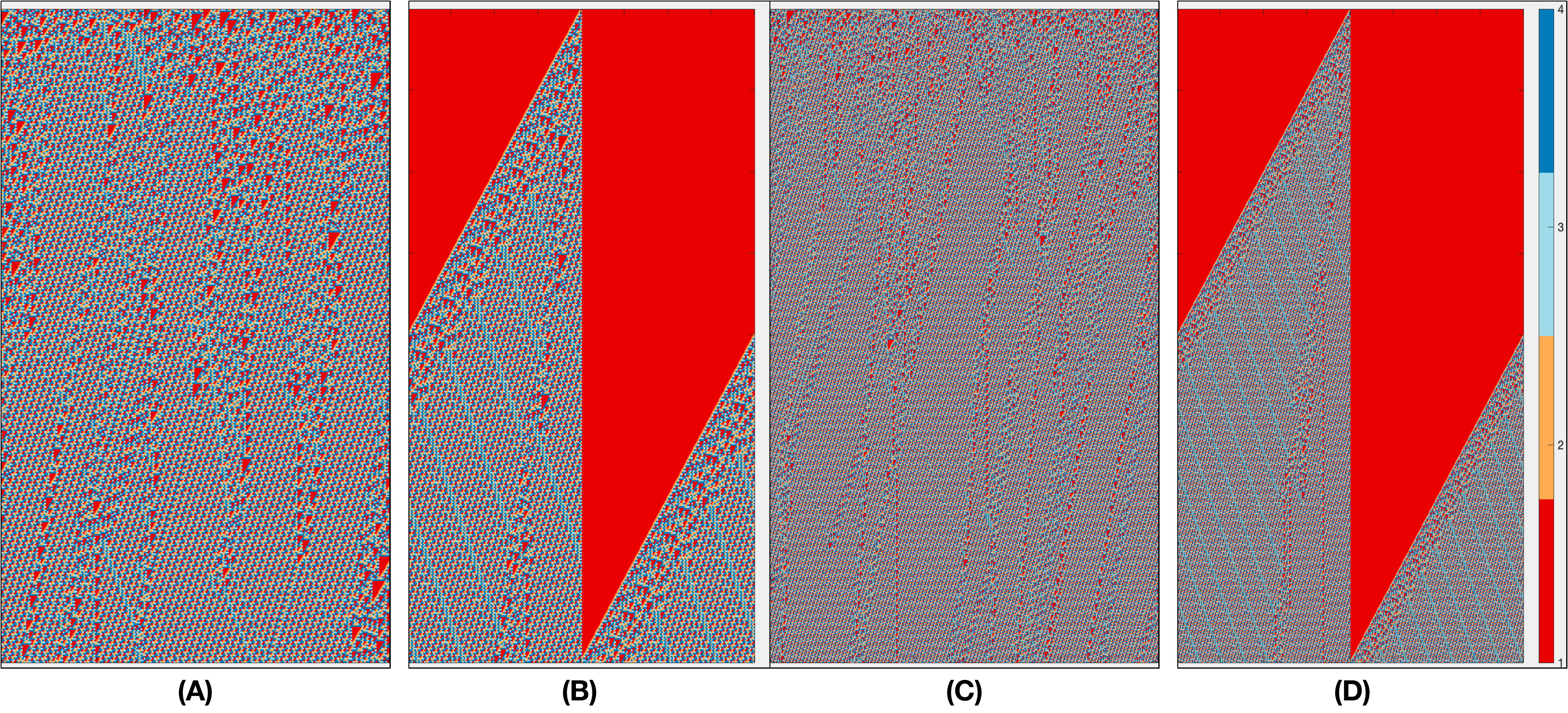}
	\caption{Evolutions of Rule 110 simulated with $4$ states.}
	\label{fig:01-evoluciones110-4h}
\end{figure}

Evolution examples of this rule are presented in Fig. \ref{fig:01-evoluciones110-4h}. Part (A) shows the evolution from a random initial configuration with $200$ cells and $400$ generations. (B) starts from a unique, different state in the initial configuration. (C) shows the evolution from a random initial configuration with $400$ cells and $800$ generations. (D) starts from a unique, different state in the initial configuration with the same number of cells and generations. In all these examples, it is clear the interaction of particles in a periodic background after a few generations.

This paper proposes the following parameters to characterize complexity in cellular automata. These criteria have been selected because they can be easily calculated in computational experiments. These parameters are:

\begin{enumerate}
	\item Diversity degree of every state as row and column in $M_\varphi$.
	\item Average of experimental density for every state measured in a number of generations.
	\item Average of experimental entropy calculated in a number of generations.
\end{enumerate}

In $M_\varphi$, the diversity degree of every row is measured with the number of different states and the repetitions of each state in the row. For every row $i$ in the evolution rule $M_\varphi$, let $U_i \subseteq S$ be the set of states in row $i$ and let $F_i$ be a vector indicating the number of times that every state $s \in U_i$ appears in row $i$. This value is indicated by $F_i(s)$.

The diversity of row $i$ ($D_i$) is calculated as:

\begin{equation}
	D_i= \sum_{s \in U_i} \left( \frac{1}{|S|} \right)^{F_i(s)}
	\label{EqDiversidadEstados}
\end{equation} 

If a row $i$ has states with high frequencies, these appear many times at row $i$, and they have a pondered value close to $0$.  On the contrary, if row $i$ has many states with few repetitions, they have low frequencies and pondered values close to $(1/|S|)$. Rows with $D$ values close to $1$ generate an extensive range of states, and rows with $D$ values close to $0$ produce a low diversity of states. This parameter is applied analogously to the columns of $M_\varphi$. Thus, we have a measure for the diversity of states produced by every state (as a row or column) of $M_\varphi$. In complex cellular automata, we are going to look for evolution rules with a mixture of states with high and low diversity degrees, which can be used to produce particles evolving in a periodic background.


Another parameter used to characterize complexity is the average density of states measured during the evolution of a cellular automaton in various computational experiments. In particular, states defining a periodic background have a specific density and states delimiting particles have a different particular density. This mixture of non-trivial densities is used to identify complex behaviors. 


The last parameter employed to determine complexity is the information entropy, measured for each state $i \in S$ in every configuration $c_j$ during the evolution of a cellular automaton. Let $|c_j|$ be the number of cells in configuration $c_j$ and let $r(i,c_j)$ be the number of cells with state $i$ in $c_j$. The entropy $E_{c_j}$ of configuration $c_j$ is calculated following Eq. \ref{EqEntropia}. 
	
\begin{equation}
E_{c_j}=-\sum_{i \in S} \left(\frac{r(i,c_j)}{|c_j|}\right) \log_{2}\left(\frac{r(i,c_j)}{|c_j|}\right)
\label{EqEntropia}
\end{equation}

This paper employs the average of information entropy evaluated in some computational experiments. Complex cellular automata have an asymptotic (not fixed or periodic) entropy behavior, indicating the average of bits needed to keep the information of every cell in each configuration.

\begin{figure}[th]
	\centering
	\includegraphics[width=0.7 \linewidth]{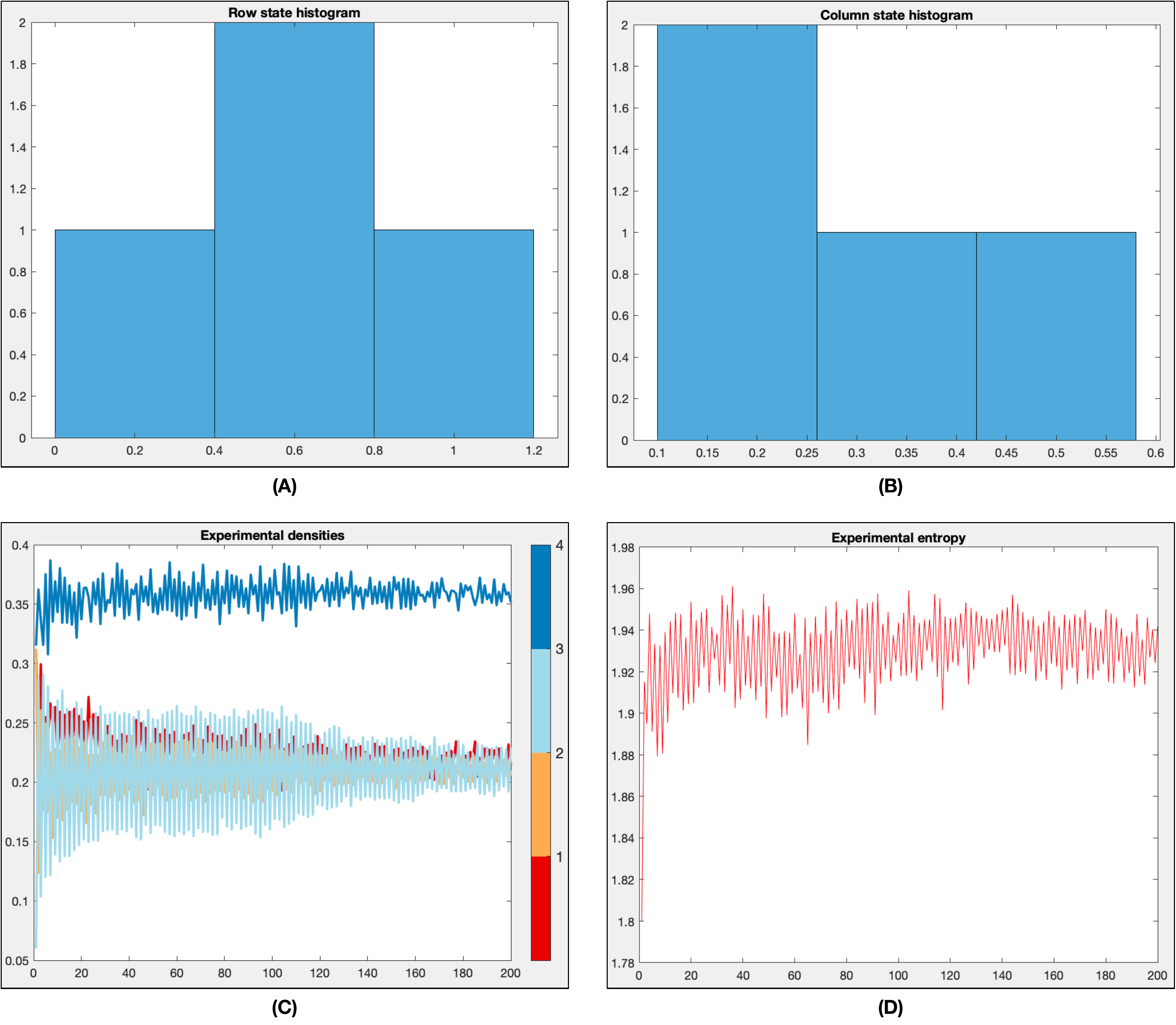}
	\caption{Properties of Rule 110 (diversity of row and column states, experimental densities and entropy).}
	\label{fig:04-propiedadescomplejas-110-4h}
\end{figure}

Figure \ref{fig:04-propiedadescomplejas-110-4h} shows histograms to classify the diversity degrees of rows and columns in $M_{110}$ (parts (A) and (B)), experimental densities and entropy of Rule 110. Histograms depict that rows have more diversity than columns, this unbalance means that states as left neighbors may evolve in states different when they act as right neighbors. This feature allows the automaton to produce a mixture of constructions generating a complex behavior.

Experimental densities (part C) have been calculated taking the average of $10$ samples with $200$ cells and $200$ generations from random initial configurations. These densities are well differentiated in two groups, showing that state $4$ is more common than the others due to the reproduction of a periodic background and the propagation of particles.

Finally, entropy (part D) has been calculated taking the average of every configuration in $10$ samples with $200$ cells and $200$ generations as well. In this case, entropy is around to $1.93$, but it is not stationary. This value indicates the interaction of particles in a periodic background, inducing a minimum but perceptible change in the dynamics of entropy in the automaton.


\section{Reversible cellular automata}

Reversible cellular automata (RCAs) are a particular type of systems where global information is conserved during the temporal dynamics. An RCA $\mathcal{A}$ has an evolution rule $\varphi$ such that there exists another inverse rule $\varphi^{-1}$ (perhaps with a different neighborhood size) inducing an invertible global mapping. It is clear that $\varphi$ and $\varphi^{-1}$ can be simulated simultaneously by evolution rules with neighborhood size $2$ \cite{Seck-2017}. Therefore, this paper only analyzes RCAs with a neighborhood size of $2$ in both rules. RCAs have been widely studied since the seminal article by Hedlund \cite{Hedlund1969}. The local properties of an RCA $\mathcal{A}$ with both invertible rules with neighborhood size $2$ can be resumed as:

\begin{itemize}
	\item Every sequence $w \in S^*$ has $|S|$ preimages.
	\item For $m >= 2$, the preimages of every $w \in S^{m}$ have a set $L_w$ of initial states with $|L_w|=L$, a common central part and a set $R_w$ of final states with $|R_w|=R$, such that $LR=|S|$.
	\item For each distinct $w,w' \in S^m$, it is fulfilled that $|L_w|=|L_w'|=L$, $|R_w|=|R_w'|=R$, $|L_w \cap R_{w'}|=1$ and $|R_w \cap L_{w'}|=1$.
\end{itemize}

Values $L$ and $R$ are known as the Welch indices of $\mathcal{A}$. These properties have been used in \cite{Seck-2017} to propose an algorithm to generate random RCAs, specifying the numbers of states and one of the Welch indices. This algorithm has been used in the experimental results of this paper to create complexity from RCAs. 

\begin{figure}
	\centering
	\includegraphics[width=1\linewidth]{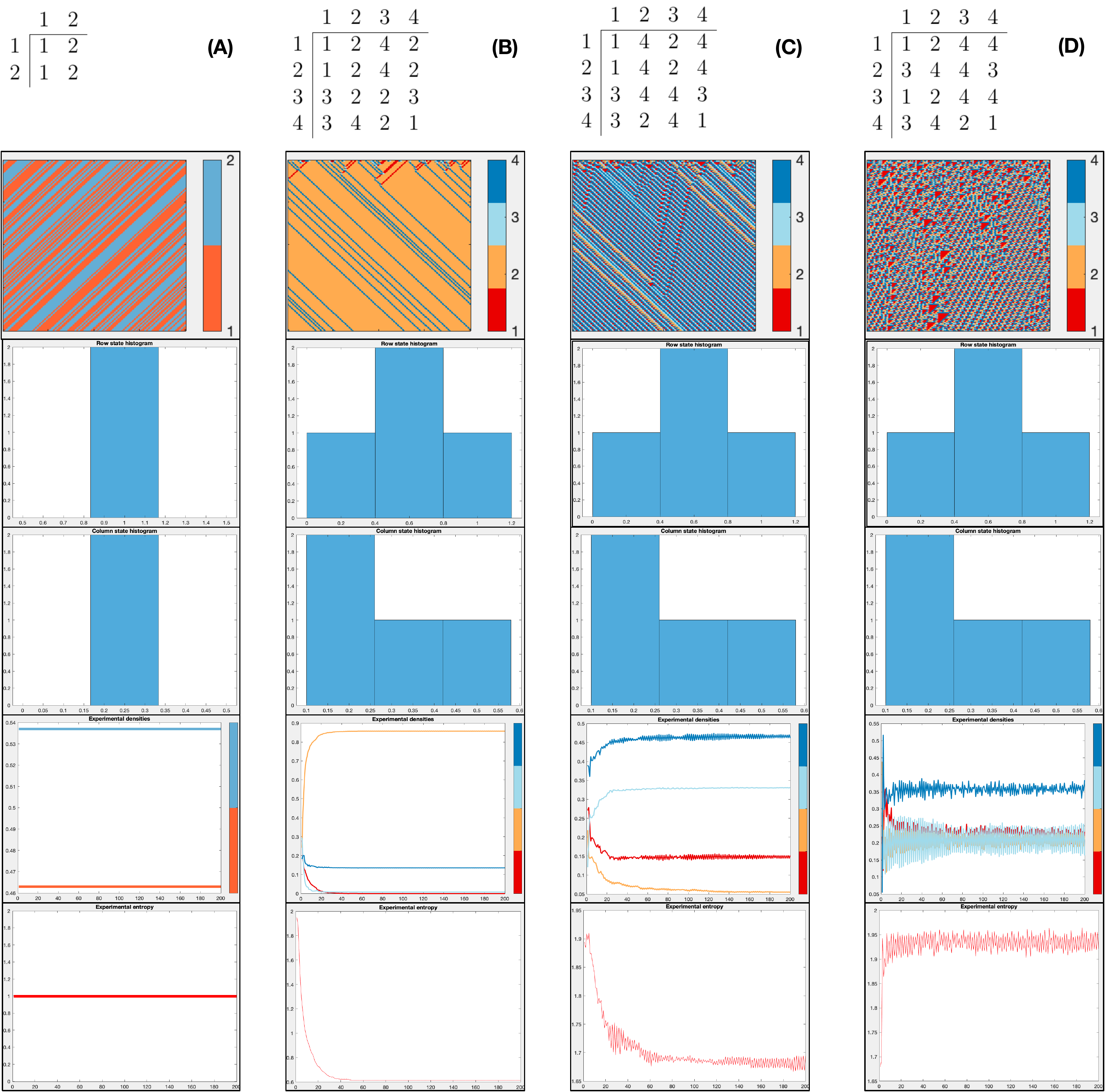}
	\caption{Transition from a reversible cellular automaton of two states to a complex cellular automaton by adding random states and permutations. State diversity, densities, and entropy are also depicted.}
	\label{fig:02-reversible-110-4h}
\end{figure}

The evolution rule in Eq. \ref{Regla110-4h} can be obtained from a reversible cellular automaton of two states and Welch index $L=2$, as shown in Fig. \ref{fig:02-reversible-110-4h}-(A). 

Evolutions in Fig. \ref{fig:02-reversible-110-4h} have $200$ cells and $200$ generations. The reversible automaton in (A) is just a right shift; more states can be added at random to obtain a new automaton with the dynamics in (B), where the shift behavior is conserved by states $2$ and $3$. Part (C) in the figure is obtained permuting states $2$ and $3$; in this evolution, we can notice the production of particles in a periodic background. Part (D) in the figure is produced by applying the same permutation  $(2 3)$  of rows and columns. In this way, the last evolution shows the complexity equivalent to Rule 110. This emergence of complex behavior is also illustrated with the histograms of state diversity,  densities, and entropy calculated in numerical experiments with $10$ samples.



\section{Complex cellular automata from reversible specimens}

\begin{figure}[th]
	\centering
	\includegraphics[width=1\linewidth]{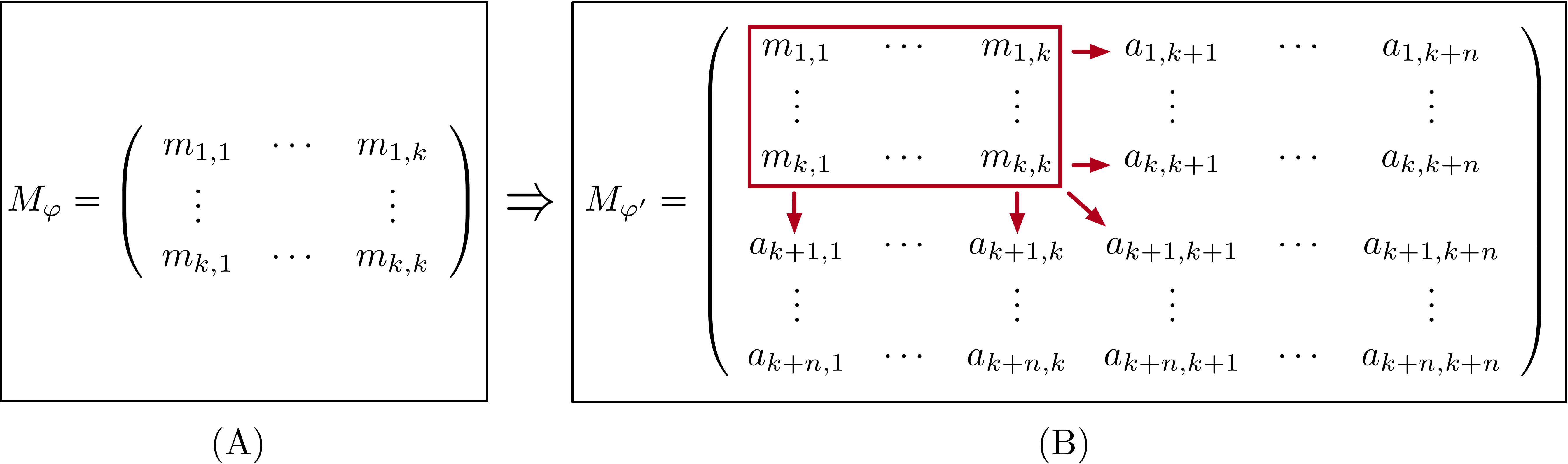}
	\caption{Extending an RCA to establish a complex cellular automata.}
	\label{fig:15-reversible-complejo-regla}
\end{figure}

In our definition of complexity, we are looking for automata where there is a set of particles interacting in a periodic background. The idea of this work is to use RCAs as a framework to establish a periodic background and then add additional states at random to obtain complexity. The process consists of generating an initial RCA with $k$ states with the algorithm presented in \cite{Seck-2017}. This algorithm can yield RCAs with dozens of states. After the RCA has been generated, an extended evolution rule is defined with $n$ additional states. The new neighborhoods are outlined at random with uniform probability, using the whole set of $k+n$ states. This process is illustrated in Fig. \ref{fig:15-reversible-complejo-regla}, a RCA with $k$ states is defined in part (A); after that, $n$ extra states are added in part (B) and the new neighborhoods are specified at random taking the $k+n$ states.

\begin{figure}[th]
	\centering
	\includegraphics[width=1\linewidth]{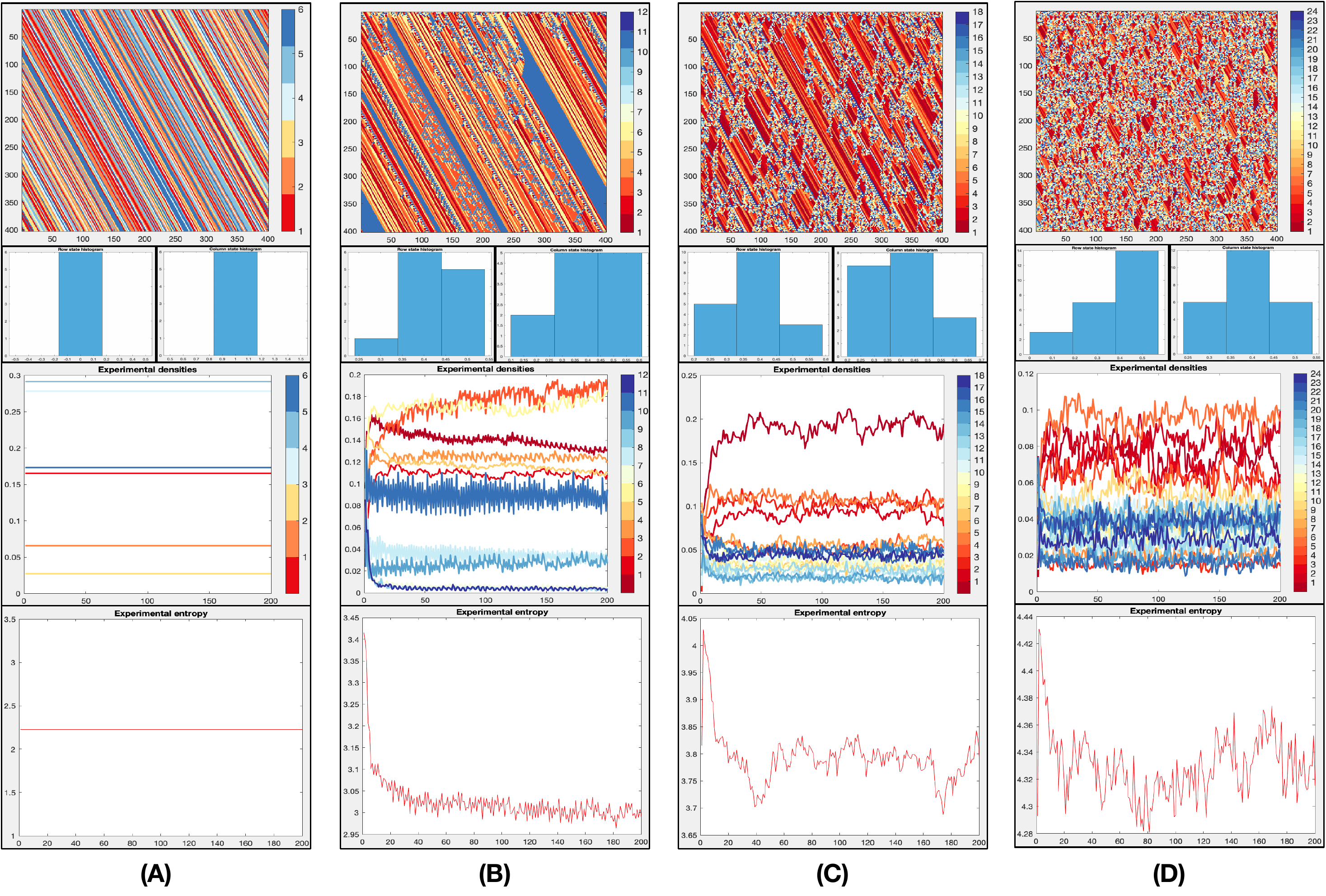}
	\caption{Obtaining complex behaviors from an RCA with $6$ states and Welch index $L=1$.}
	\label{fig:05-reversible-complejo-6hw1}
\end{figure}

Figure \ref{fig:05-reversible-complejo-6hw1} presents an instance of this process; first a trivial RCA with $6$ states and Welch index $L=1$ (a right shift) is defined (part (A)); we can notice that all states have the same diversity measure classified in histograms below the temporal evolution sample of $200$ cells and $200$ generations. Experimental densities below histograms are conserved during the whole evolution, as we can expect from an RCA. Finally, the experimental entropy shows that the information is uniformly conserved during the dynamical behavior of the automaton. 

Random states were added to the evolution rule to obtain a cellular automaton with $12$ states (part(B)), temporal evolution presents the rise of some particle-like structures, but the periodic background is still too dominant. Histograms present more diversity in the evolution of every state, but experimental densities are still too close, and entropy is around $3$ with small deviation as the automaton evolves. When $18$ states are considered ($12$ random additional states in part (C)), we have a mixture of particles interacting in a periodic background. Histograms show more state diversity; as rows, average diversity is predominant and as columns low and average diversity are similar. Densities can be well classified in three groups, indicating that some states are more familiar to define the periodic background, and other states are used only in the particle formation. Entropy has a more interesting behavior, centered around $3.75$ with small fluctuations close to $0.05$ at both directions.

Last part of Fig. \ref{fig:05-reversible-complejo-6hw1} displays the case when $18$ additional random states were considered. Histograms tend to average and high diversities, densities are almost merged indicating that states are almost equiprobable, and entropy is more unstable around $4.33$, and with a deviation close to $0.4$, this indicates that a chaotic behavior has been reached.


Computational experiments indicate that it is more common to obtain complexity when additional random states are considered approximately a proportion of $1:2$; that is, $k$ states defining the RCA and around $2k$ random states completing the evolution rule. This proportion yields a mixture between the emergence of states in the original RCA which act in the specification of a periodic background, and additional states producing particles interacting in this environment.

A set of experiments has been done to observe the rate of complex behaviors obtained when RCAs are expanded at random. Table \ref{Tabla-PorcentajeComplejos} shows different cases of RCAs with distinct Welch indices. The first row shows the different cases $(k,k+n)$ where $k$ is the number of states defining an RCA, and $k+n$ is the total number of states in the final automaton. For every case, $100$ different evolution rules were generated and analyzed using diversity in the evolution rule, the density of states and entropy, to obtain a percentage of complex rules obtained with this process. These experiments have been defined with $400$ cells and $800$ evolutions to facilitate the computational study.

\begin{table}[th]
	\centering
\begin{tabular}{|c|c|c|c|c|c|}
	\hline 
	Size & $(6,18)$ & $(7,21)$ & $(16,48)$ & $(18,54)$ & $(19,57)$ \\ 
	\hline 
	$L=1$ & $22\%$ & $15\%$ & $21\%$ & $20\%$ & $20\%$ \\ 
	\hline 
	$L=2$ & $13\%$ &  & $17\%$ & $12\%$ &  \\ 
	\hline 
	$L=3$ &  &  &  & $8\%$ &  \\ 
	\hline 
	$L=4$ &  &  & $13\%$ &  &  \\ 
	\hline 
\end{tabular} 
\caption{\label{Tabla-PorcentajeComplejos}Percentage of complex rules obtained from different RCAs.}
\end{table}

There are blanks in Table \ref{Tabla-PorcentajeComplejos} because not every Welch index divides the number of states of each RCA. This table shows that complex behaviors are not strange even when more states are considered, but complexity is more difficult to be obtained when higher values of Welch indices are considered.     

\begin{figure}[th]
	\centering
	\includegraphics[width=\linewidth,height=0.69 \linewidth]{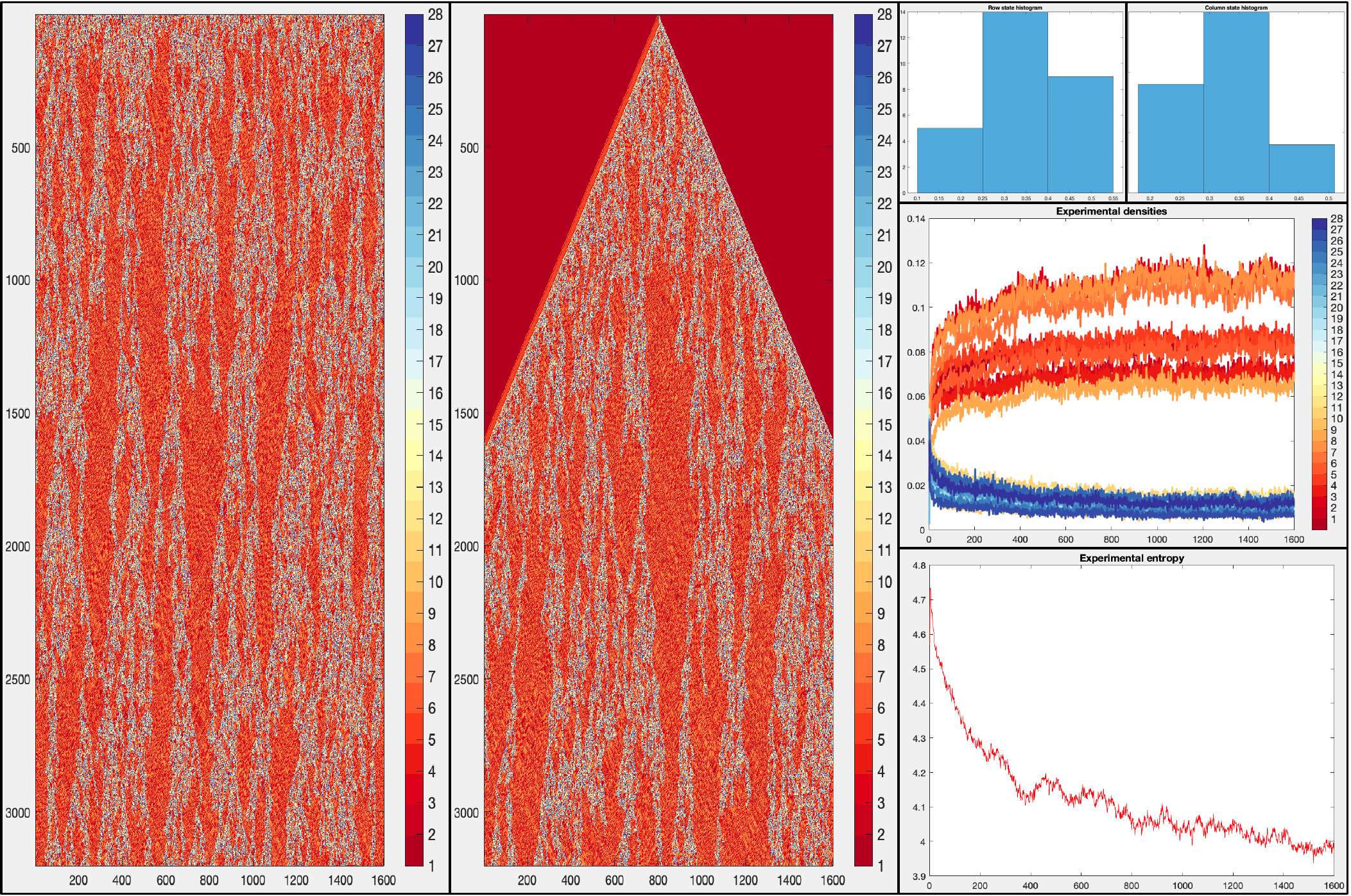}
	\caption{Complex cellular automaton of $28$ states from an RCA with $9$ states and Welch index $L=3$.}
	\label{fig:06-EjemplosComplejos-1}
\end{figure}
\begin{figure}[th]
	\centering
	\includegraphics[width=1\linewidth,height=0.87 \linewidth]{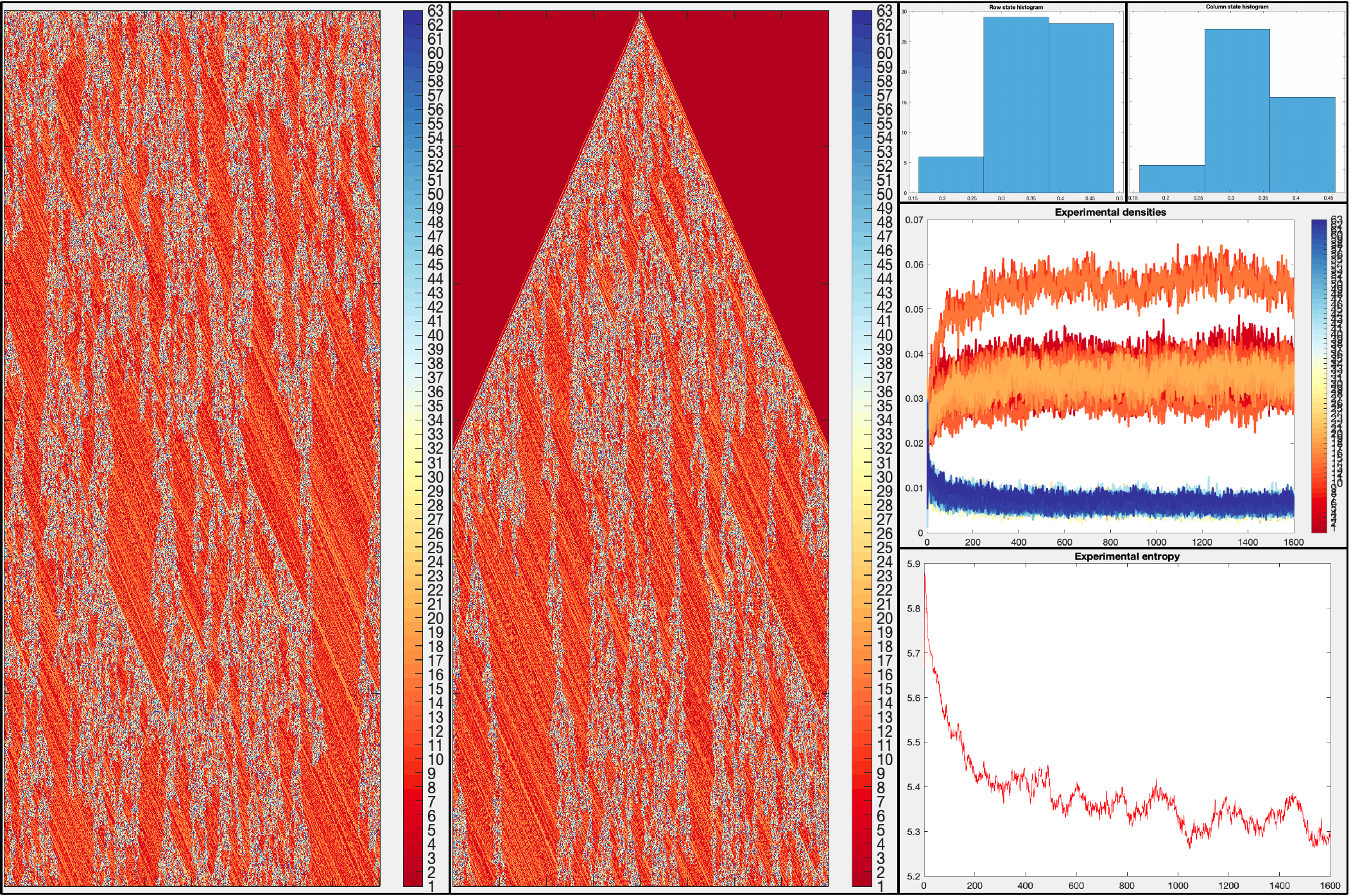}
	\caption{Complex cellular automaton of $63$ states from an RCA with $20$ states and Welch index $L=2$.}
	\label{fig:07-EjemplosComplejos-2}
\end{figure}
\begin{figure}
	\centering
	\includegraphics[width=1\linewidth,height=0.87 \linewidth]{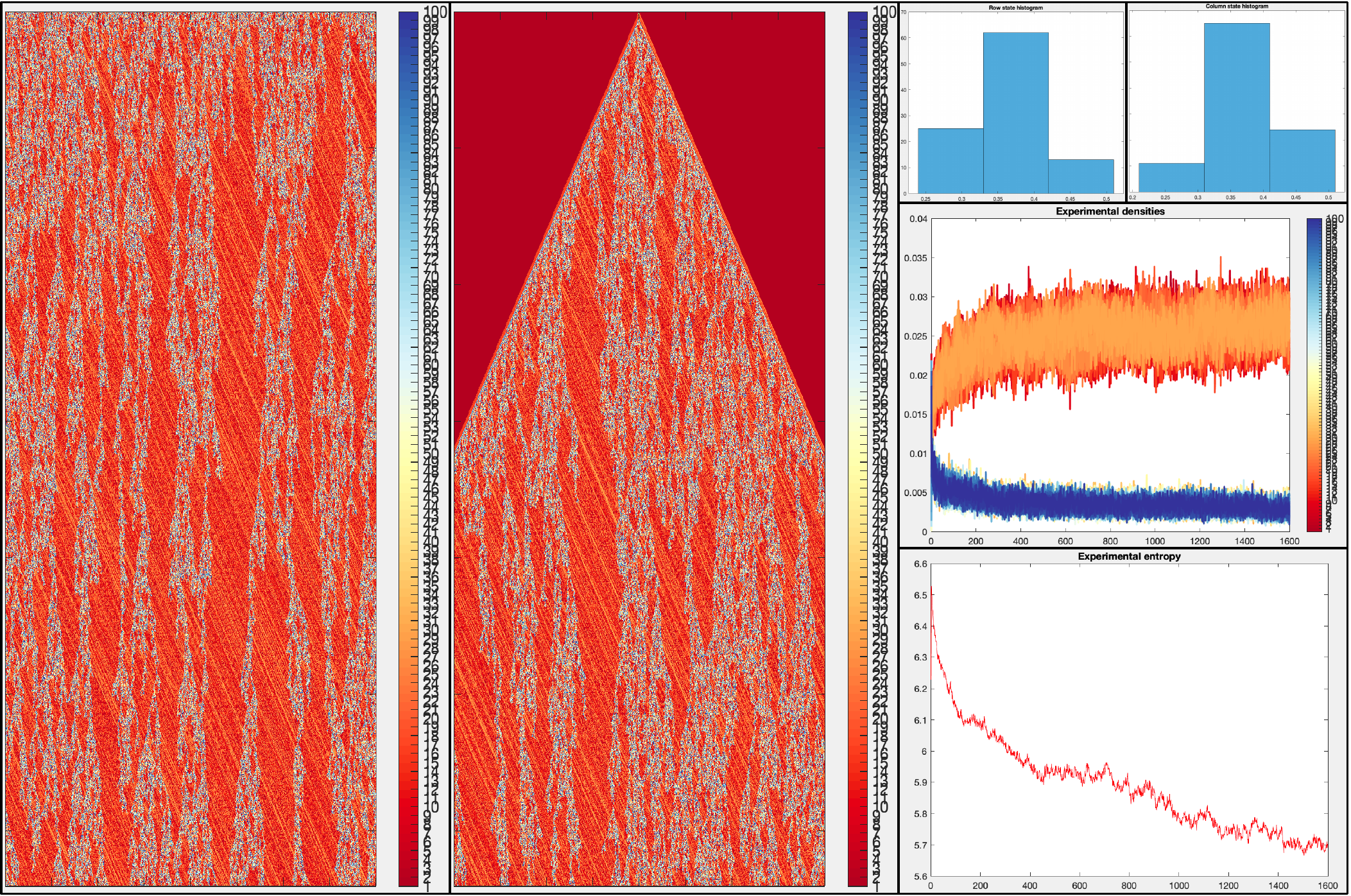}
	\caption{Complex cellular automaton of $100$ states from an RCA with $30$ states and Welch index $L=3$.}
	\label{fig:08-EjemplosComplejos-3}
\end{figure}

Figures \ref{fig:06-EjemplosComplejos-1}, \ref{fig:07-EjemplosComplejos-2} and \ref{fig:08-EjemplosComplejos-3} present examples of some complex cellular automata generated from RCAs. All cases have $800$ cells and $1600$ generations. Every example has two evolutions, one generated from a random configuration, and another from a fixed state. Histograms classifying the diversity of states are also depicted, and the average of state density and entropy are displayed as well.

\newpage

Figure \ref{fig:06-EjemplosComplejos-1} shows a complex cellular automaton with $28$ states generated from a RCA with $9$ states and Welch index $L=3$. There is a mixture in the diversity of states as rows or columns in the evolution rule. Densities describe $3$ groups of different values and the entropy is asymptotically decreasing with fluctuations.

Figure \ref{fig:07-EjemplosComplejos-2} illustrates a complex cellular automaton with $63$ states generated from a RCA with $20$ states and Welch index $L=2$. Again, there is a combination in the histograms depicting the diversity of states. Densities are grouped in $3$ groups, and the entropy is asymptotically decreasing as well.

\newpage

Figure \ref{fig:08-EjemplosComplejos-3} presents a complex cellular automaton with $100$ states generated from a RCA with $30$ states and Welch index $L=3$. Histograms have a mixture of diversity, and now densities are grouped in $2$ classes, entropy is asymptotically decreasing with fluctuations.

In the previous experiments, a concrete value for entropy has not been characterized when complexity is generated. However, these experiments show that entropy is close and always below $log(k+n)$ to obtain complex behaviors.

\section{Variation of the reversible part by permutations}


The algorithm used in this paper is characterized to produce reversible automata with quiescent states; that is, for the reversible evolution rule $\varphi$ generated by the algorithm, the matrix $M_\varphi$ fulfills that every diagonal element $M_\varphi(i, i)=i$. There are transformations over $M_\varphi$ which conserve reversibility; in particular, the effect of permutation of states and permutation of row and columns are illustrated in this paper.

For a RCA of $k$ states with evolution rule $\varphi$ represented by $M_\varphi$, and a permutation $\pi:\{1 \ldots k \} \rightarrow \{1 \ldots k \} $, a permutation of states over $M_\varphi$ defines a new RCA with rule $\varphi'$ such that $M_{\varphi'}(i,j)=\pi(M_\varphi(i,j))$ for $1 \leq i \leq j \leq k$. In a similar way, a permutation of rows and columns over $M_\varphi$ defines a new RCA with rule $\varphi'$ such that $M_{\varphi'}(\pi(i),\pi(j))=M_\varphi(i,j)$. 

\begin{figure}[th]
	\centering
	\includegraphics[width=1\linewidth,height=0.945 \linewidth]{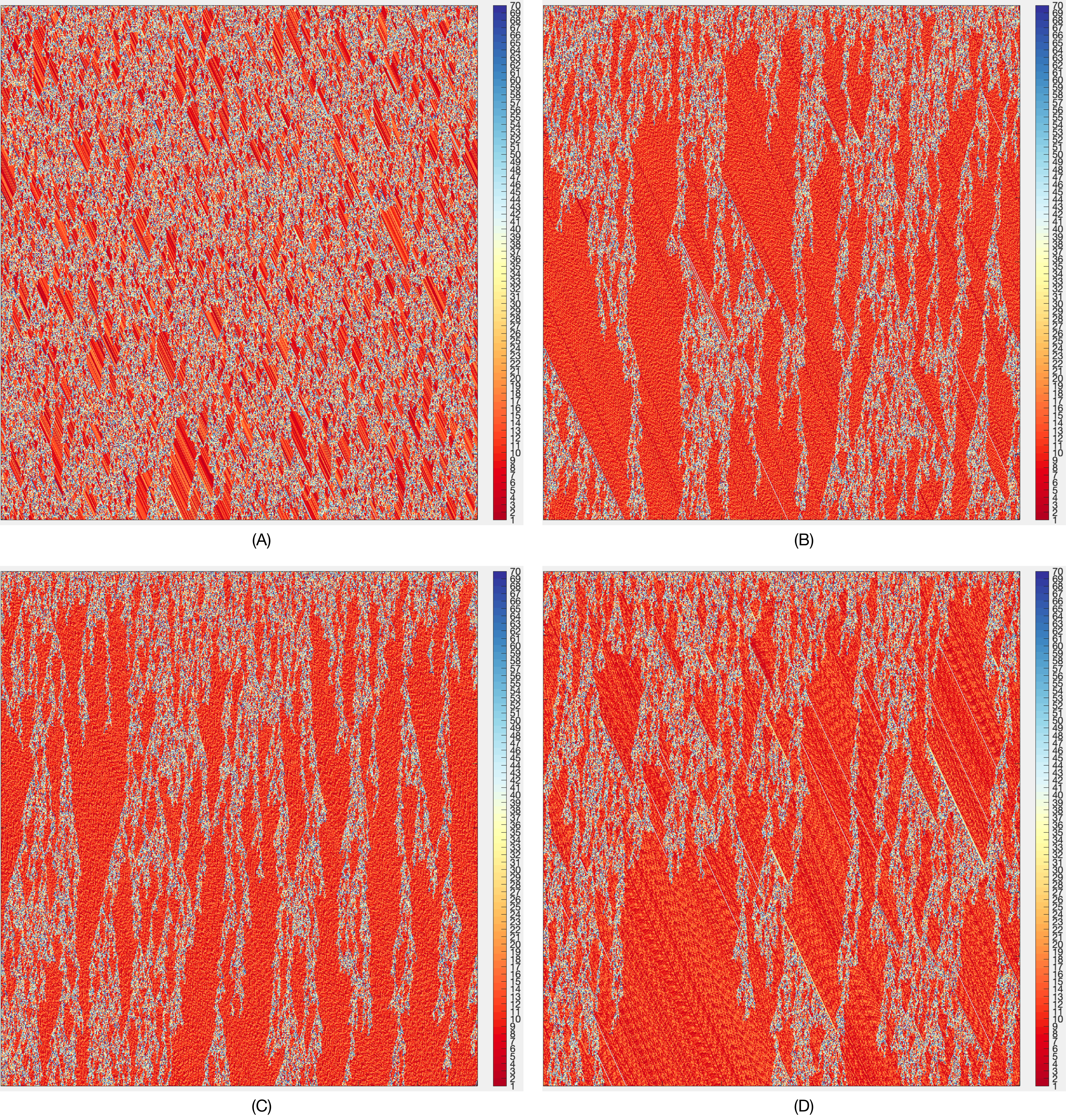}
	\caption{Complex cellular automata of $70$ states generated from an RCA with $20$ states, Welch index $L=1$ and applying permutation of states and permutations of rows and columns.}
	\label{fig:16-Reversible-Complejo-Permutaciones-70-20-1-800-800}
\end{figure}

\begin{figure}[th]
	\centering
	\includegraphics[width=1\linewidth,height=0.945 \linewidth]{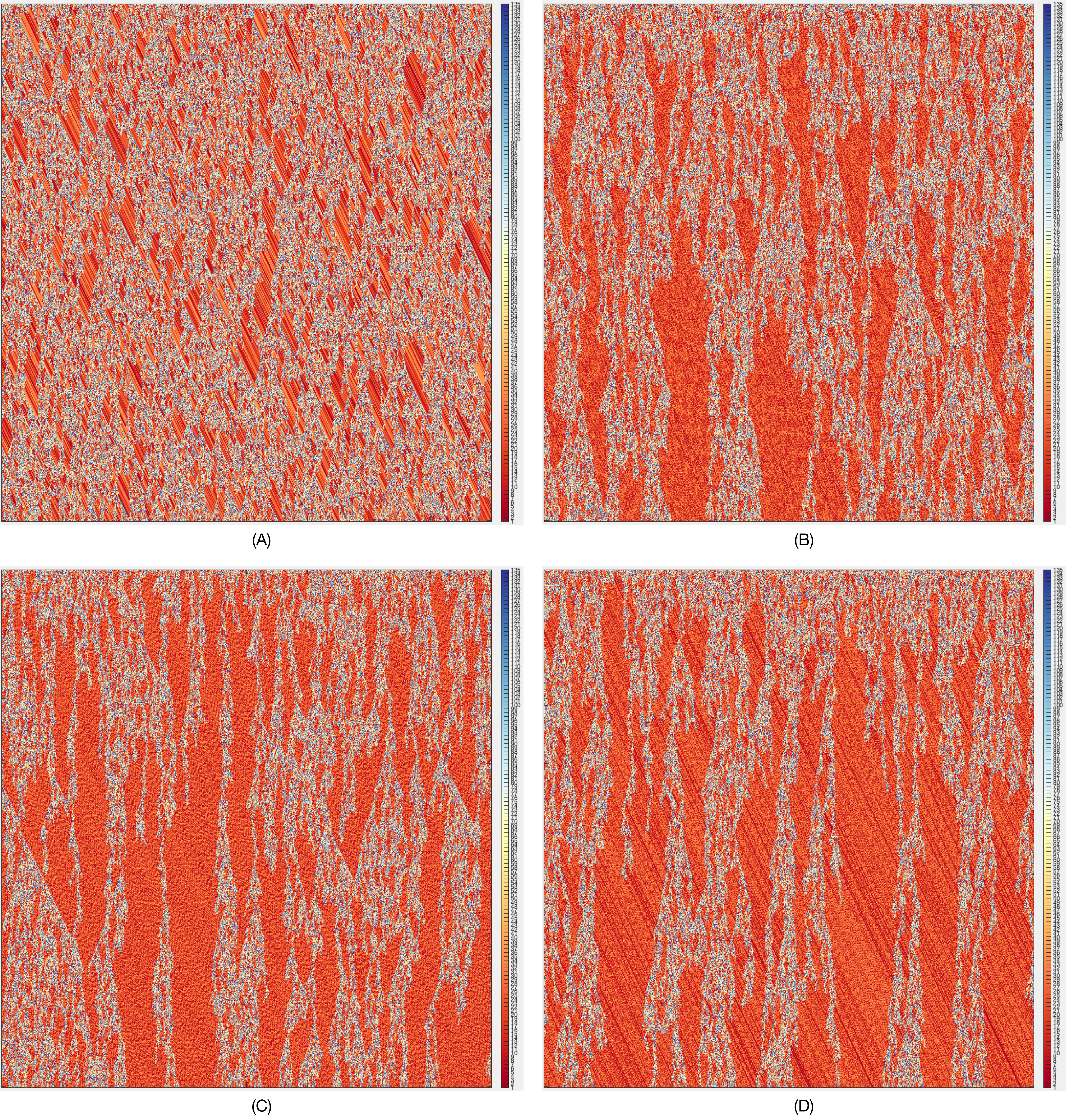}
	\caption{Complex cellular automata of $135$ states generated from an RCA with $40$ states, Welch index $L=1$ and applying permutation of states and permutations of rows and columns.}
	\label{fig:17-Reversible-Complejo-Permutaciones-135-40-1-800-800}
\end{figure}

Figures \ref{fig:16-Reversible-Complejo-Permutaciones-70-20-1-800-800} and \ref{fig:17-Reversible-Complejo-Permutaciones-135-40-1-800-800}  are instances of some complex cellular automata generated from RCAs with left Welch index $L=1$. These cases have $800$ cells and $800$ generations. Figure \ref{fig:16-Reversible-Complejo-Permutaciones-70-20-1-800-800} shows some cellular automata with $70$ states generated from a RCA with $20$ states and Welch index $L=1$. Part (A) is the original RCA with quiescent states, part (B) shows a modification obtained by a permutation of states. Part (C) depicts another change using a permutation of rows and columns, and part (D) has been obtained applying both types of permutations at the same time. The original automaton is almost dominated by chaotic behavior; only some isolated parts are visible with a periodic structure derived from the RCA. However, when a state permutation is applied over the reversible part, the dynamical behavior is closer to complexity. The same is observed in (C) and (D).  Figure \ref{fig:17-Reversible-Complejo-Permutaciones-135-40-1-800-800} illustrates some cellular automata with $135$ states generated from a RCA with $40$ states and Welch index $L=1$. Part (A) is based on the original RCA, part (B) uses a permutation of states, part (C) applies a permutation of rows and columns, and part (D) is a mixture of both permutations.

These examples demonstrate that complexity can be obtained even when dozens of states are considered, as long as some of the states determine a stable dynamical structure; in the previous cases, reversibility brings an adequate framework to define particles in a periodic background. Meanwhile, the original RCAs are close to chaos, permutations in states, rows, and columns provide a stronger structure to the reversible part, provoking a more complex dynamical behavior.  
 
\newpage

\newpage

\newpage

\begin{figure}[th]
	\centering
	\includegraphics[width=1\linewidth,height=0.945 \linewidth]{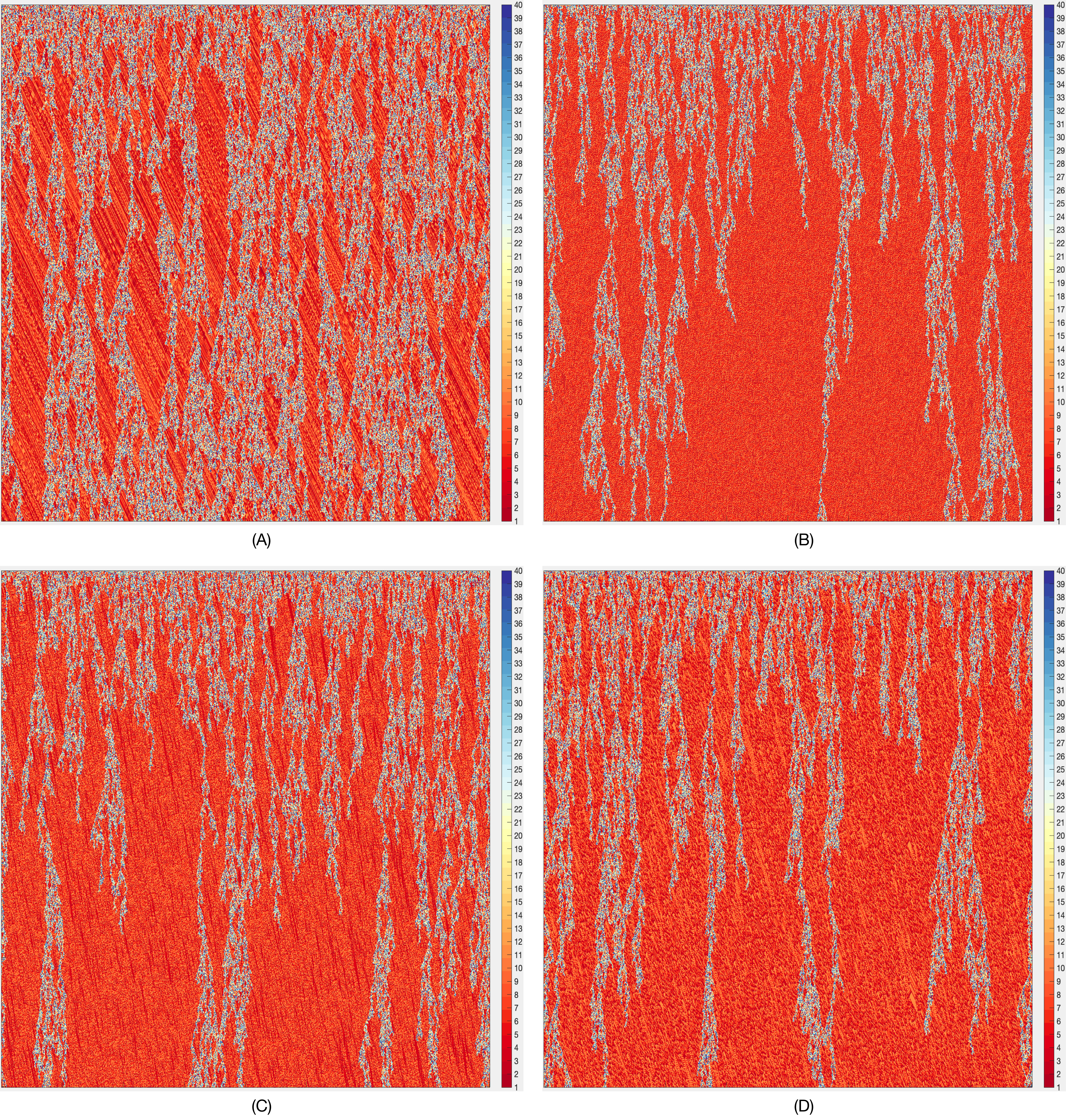}
	\caption{Complex cellular automata of $40$ states generated from an RCA with $12$ states, Welch index $L=2$ and applying permutation of states and permutations of rows and columns.}
	\label{fig:18-Reversible-Complejo-Permutaciones-40-12-2-800-800}
\end{figure}

\begin{figure}[th]
	\centering
	\includegraphics[width=1\linewidth,height=0.945 \linewidth]{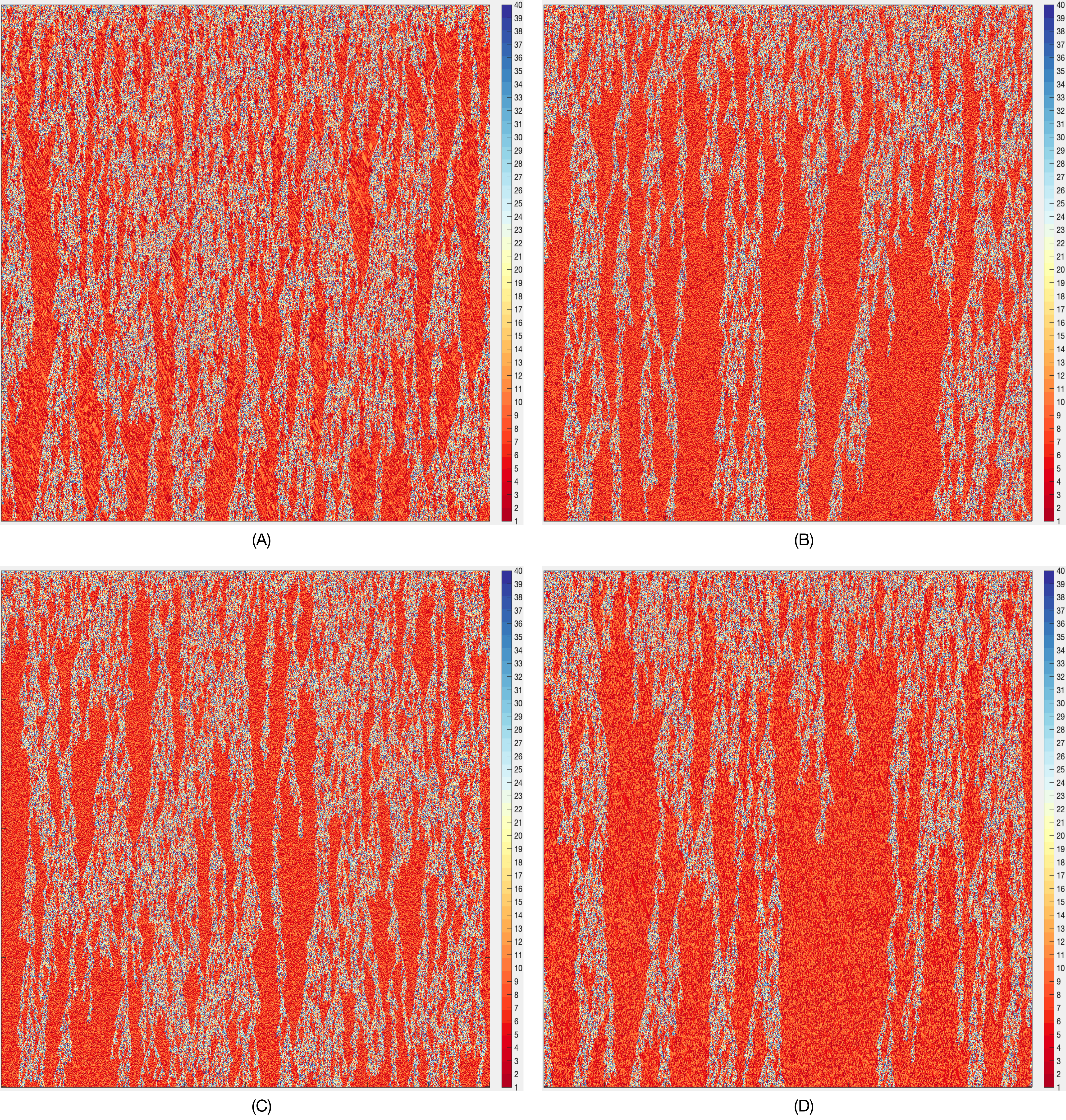}
	\caption{Complex cellular automata of $40$ states generated from an RCA with $12$ states, Welch index $L=3$ and applying permutation of states and permutations of rows and columns.}
	\label{fig:19-Reversible-Complejo-Permutaciones-40-12-3-800-800}
\end{figure}

\begin{figure}[th]
	\centering
	\includegraphics[width=1\linewidth,height=0.945 \linewidth]{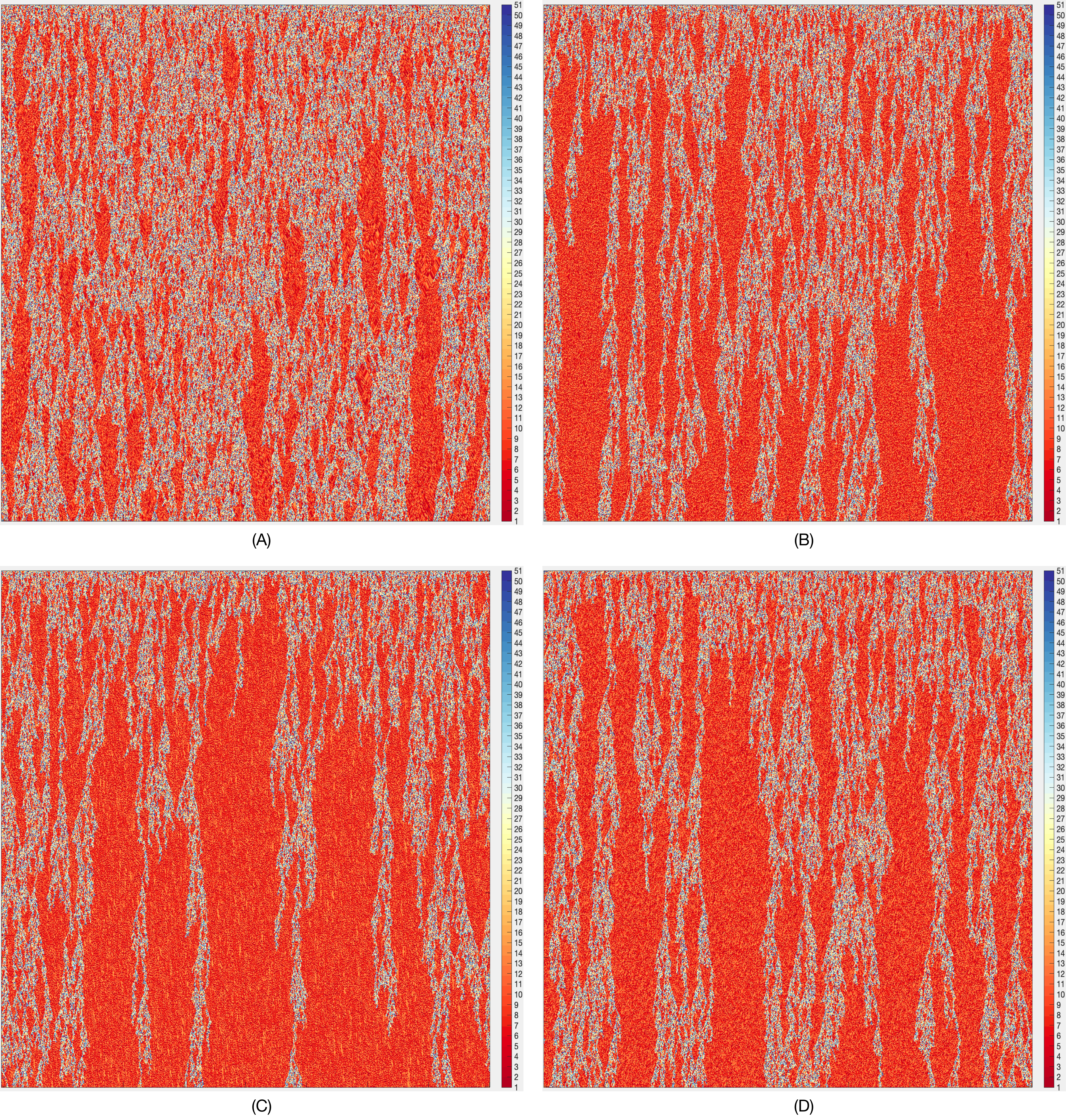}
	\caption{Complex cellular automata of $51$ states generated from an RCA with $15$ states, Welch index $L=3$ and applying permutation of states and permutations of rows and columns.}
	\label{fig:20-Reversible-Complejo-Permutaciones-51-15-3-800-800}
\end{figure}

The previous examples are constructed with RCAs having a unitary Welch index. It is interesting to investigate if the same dynamical behavior can be obtained with RCAs with nonunitary Welch indices. Figures  	\ref{fig:18-Reversible-Complejo-Permutaciones-40-12-2-800-800}, 	\ref{fig:19-Reversible-Complejo-Permutaciones-40-12-3-800-800} and \ref{fig:20-Reversible-Complejo-Permutaciones-51-15-3-800-800} present instances of complex cellular automata generated from RCAs with nonunitary Welch indices. 

\newpage

\newpage

The previous figures depict cellular automata with $40$ and $51$ states produced from an RCA with $12$ and $15$ states and Welch indices $L=2$ and $L=3$. Part (A) is the original RCA with quiescent states and dynamical behavior closer to chaos, part (B) shows a modification obtained by a permutation of states. Part (C) presents a permutation of rows and columns, and part (D) has been obtained with both types of permutations. Again, when some permutation is applied over the reversible part, the dynamical behavior is closer to complexity, with particles interacting in a periodic background. 


\section{Discussion}

\begin{figure}[th]
	\centering
	\includegraphics[width=1\linewidth,height=0.98 \linewidth]{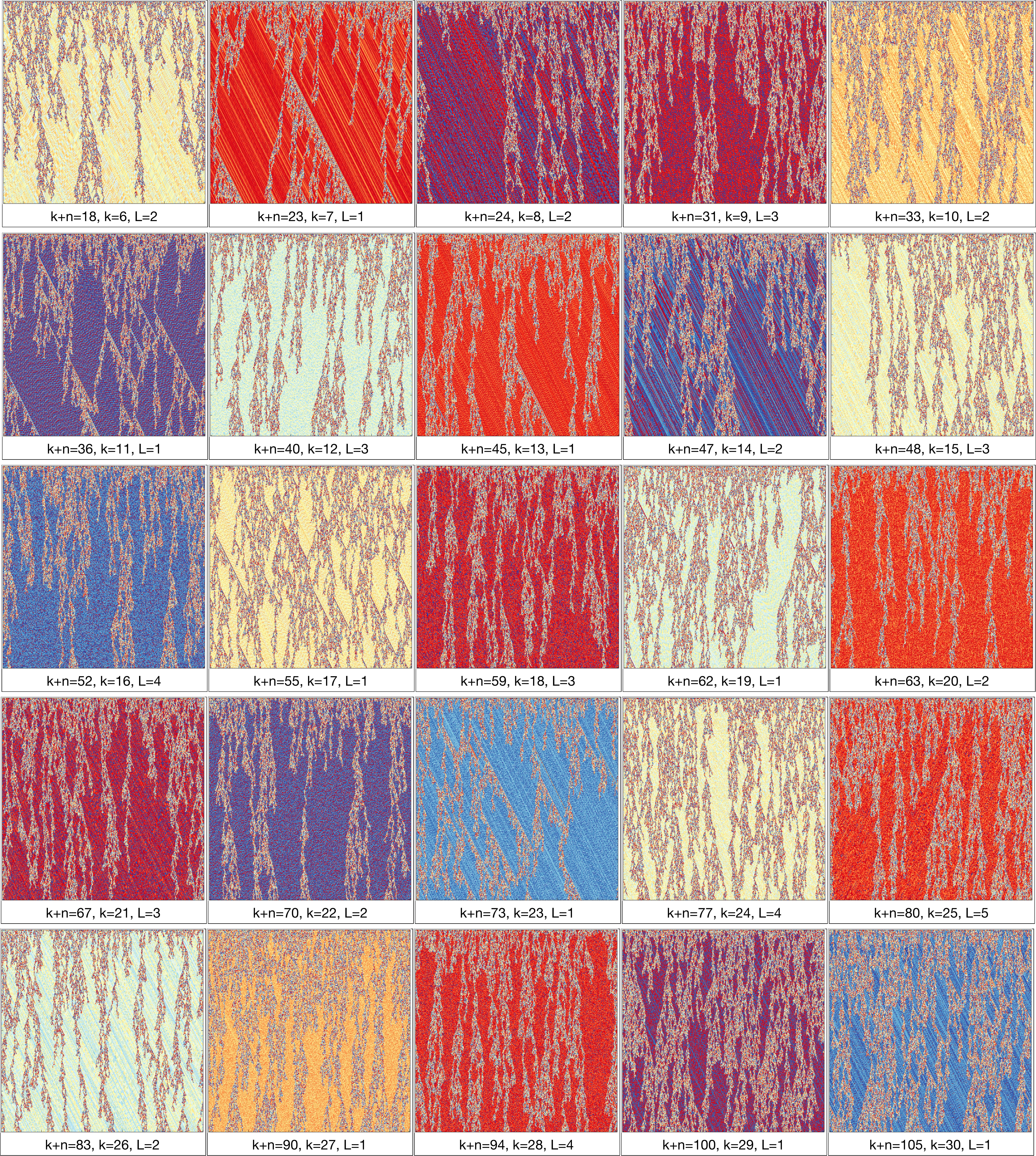}
	\caption{Examples of complex cellular automata from $18$ to $105$ states obtained from RCAs with a mixture of Welch indices.}
	\label{fig:21-Collage-Reversibles-500-1000}
\end{figure}

Experimental results demonstrate that a proportion close to $(3,1)$ (number of states, number of reversible ones) is adequate to obtain complex behaviors from RCAs, characterized by a diversity of states, entropy, and density of states. Meanwhile, the number of states in the original RCA is not determinant to obtain complexity and Welch indices have more influence to produce complex behaviors. Nevertheless, if not general, it is not strange as well to find complexity expanding RCAs at random as it can be observed in Table \ref{Tabla-PorcentajeComplejos}. Permutations of the reversible part (states or rows and columns) show an essential effect to reduce chaos and yield complex behaviors when more states are considered.

Figure \ref{fig:21-Collage-Reversibles-500-1000} presents a collage of different complex cellular automata from $18$ to $105$ states obtained with RCAs defined from $6$ up to $30$ states and different values of Welch indices.

\section{Conclusions}

This paper has demonstrated that RCAs define an adequate framework to produce complex cellular automata when they are expanded at random. The procedures presented in this paper can produce complexity with dozens of states. Meanwhile, the number of states is not a constraint to obtain complex behaviors, Welch indices are more determinant to restrict the rise of complexity. Our experiments show a proportion close to $k$ reversible states for $3k$ total states where complexity is more probable to be obtained.

Permutations of the reversible part have proved their utility to transit from chaos to complex behavior, generating a periodic background with interacting particles even when a large number of states are considered. It is interesting to observe how a large number of states reach self-organization defining particles and background. 

Further work may imply the generation of complex cellular automata in a more intelligent or heuristic way than only expanding at random RCAs; for example, employing fuzzy or evolutionary algorithms. Besides, to prove that these cellular automata are computationally universal is also an open problem to be treated. Another direction can be the application of more complex measures and metrics to construct complex automata from RCAs. It implies more computational time or resources, but the smart construction of complex automata can be useful to reduce the number of specimens to review. An analytical method defining how to create complex automata from a given RCA would be desirable, even for a small number of states.

Besides the application of intelligent or heuristic techniques, another kind of automata can be used to support complex behaviors; for instance, surjective automata, the use of partitions from the full shift system as used in symbolic dynamics.

\section*{Acknowledgment}
This work has been supported by CONACYT project No. CB- 2017-2018-A1-S-43008 and by IPN Collaboration Network ``\,Grupo de Sistemas Complejos del IPN\,''.

\bibliography{articuloReversibles}
\bibliographystyle{plain}

\end{document}